# Heavy Flavour Production in ALICE

Rachid Guernane for the ALICE Collaboration

*Laboratoire de Physique Corpusculaire de Clermont-Ferrand, Université Blaise Pascal and CNRS-IN2P3, 24, Avenue des Landais, 63177 AUBIÈRE CEDEX FRANCE*

**Abstract.** We review the most recent studies on the performance of ALICE in heavy flavour production measurements in both hadronic and semileptonic decay channels.

## INTRODUCTION: HEAVY FLAVOUR PRODUCTION AT THE LHC

Heavy flavour production at hadron colliders allows access to a rich QCD phenomenology [1,2]. In nucleon-nucleon collisions, heavy quark-antiquark pair production should be well described with the parton-model approach since the heavy quark mass can be considered sufficiently large with respect to the fundamental QCD scale [3]. Perturbative QCD cross-sections for heavy flavour production have been computed up to Next-to-Leading-Order accuracy [4], even if quantitatively large uncertainties are expected at LHC energies, typically of a factor 2÷3 [2]. Recently, FONLL calculation [5] for $b$ production cross-section comes to agreement within errors with CDF Run II data [6]. Open heavy flavour production is considered to be a valuable probe for studying properties of the dense QCD medium formed in heavy-ion collisions [7,8]. Heavy quarks, once produced in the early stages of the collision, are expected to rescatter and lose energy traversing the surrounding matter, testing this way the expected mass dependence of parton energy loss [9]. LHC benchmark cross-sections computed using a Fixed-Order (NLO) massive calculation from Mangano *et al.* are presented in Tab. 1 [10].

**TABLE 1.** ALICE benchmark for total heavy quark production cross-sections in p-p and central Pb-Pb collisions. For Pb-Pb, PDF nuclear modification ($C_{\text{shad}}$ being defined as the ratio of the nucleon-nucleon cross-sections $\sigma^{NN}_{Q\bar{Q}}$ calculating with and without shadowing effects from the EKS98 parametrization [11]) is included and a binary scaling from nucleon-nucleon cross-sections assumed.

| **Colliding system** | **Pb-Pb** | | **p-p** | |
|---|---|---|---|---|
| $\sqrt{s_{NN}}$ **[TeV]** | 5.5 | | 14 | |
| **Centrality** | $0 \div 5\,\%\,\sigma^{\text{inel}}$ | | | |
| **Flavour** | $c\bar{c}$ | $b\bar{b}$ | $c\bar{c}$ | $b\bar{b}$ |
| $\sigma^{NN}_{Q\bar{Q}}$ **[mb]** | 6.64 | 0.21 | 11.2 | 0.51 |
| $C_{\text{shad}}$ | 0.65 | 0.86 | | |
| $N_{Q\bar{Q}}$ **per collision** | 115 | 4.56 | 0.16 | 0.0072 |

In particular, it will be important in ALICE to measure heavy flavoured particle transverse momentum distributions which are expected to be remarkably sensitive to many nuclear effects [12]. The low-$p_t$ (below 6 GeV/c at the LHC) region is sensitive to non-perturbative effects (such as flow, quark coalescence, gluon shadowing, Color-Glass-Condensate state...); while the high-$p_t$ region is sensitive to jet quenching, namely an attenuation (or slowing down) of heavy quark jets after propagation through the dense medium.

Finally, since we consider that a $Q\overline{Q}$ pair produced in a hadronic collision may be converted into either a bound state or a pair of heavy flavoured particles[1], open heavy flavour production is regarded as the most natural baseline for quarkonia production in order to address any suppression/enhancement due to the presence of a QGP phase.

In what follows, we shall discuss the ALICE perspectives for the measurement of heavy flavour production at the LHC [13].

## ALICE SUBSYSTEMS FOR HEAVY FLAVOUR MEASUREMENTS

ALICE is a general-purpose experiment equipped with detectors capable of measuring and identifying hadrons, leptons, and photons across a large range of transverse momentum from around $100\,\mathrm{MeV}/c$ to about $100\,\mathrm{GeV}/c$ [14]. The ALICE design has been optimized to cope with the very demanding environment of high-multiplicity central Pb-Pb collisions at LHC energies, where up to 8000 charged particles per rapidity unit at mid-rapidity have been predicted. The studies presented hereafter have been carried out assuming a conservative charged particle rapidity density of $\mathrm{d}N_{\mathrm{ch}}/\mathrm{d}y = 6000$ (note that RHIC results tend to favor multiplicity values lower by about a factor 2.5 [15]). The ALICE detector has a good acceptance for heavy flavour detection. Heavy flavour decay products are reconstructed in a central barrel ($-0.9 < \eta < 0.9$) embedded in a large magnet (L3) providing a weak solenoidal field $< 0.5\,\mathrm{T}$, complemented by a forward muon spectrometer ($-4 < \eta < -2.5$). In the ALICE central barrel, charm and beauty particles are selected out of the large background, taking advantage of their long lifetime[2]. The ALICE detection strategy relies on resolving secondary detached vertices consisting of identified tracks with large impact parameters ($d_0$), the impact parameter being the distance of closest approach of a particle trajectory to the primary vertex. Precision on secondary vertex determination is provided by the two innermost layers of the inner tracking system (ITS) made of silicon pixel detectors (SPD). A resolution $\sigma_{d_0}(r\varphi) < 60\,\mu\mathrm{m}$ is achieved for $p_{\mathrm{t}} \gtrsim 1\,\mathrm{GeV}/c$. Particle tracking relies on the six concentric layers of high-resolution silicon detectors of the ITS: the two SPD layers quoted above, plus two layers of silicon drift detector (SDD), and two layers of silicon strip detector (SSD), a large volume time-projection chamber (TPC), and a high-granularity transition-radiation detector (TRD)[3]. Particle identification in the central region is performed over the full azimuth by a $\mathrm{d}E/\mathrm{d}x$ measurement in the tracking detectors, a large area high-resolution array of TOF counters, and transition radiation in the TRD. A spectrometer dedicated to muon detection and identification in the forward region complete the ALICE set-up. The ALICE muon spectrometer is made of a passive front absorber of total thickness corresponding to ten interaction lengths to absorb hadrons and photons from the interaction vertex, a high-granularity tracking system of ten planes of cathode pad chambers, a large dipole magnet creating a field of $0.7\,\mathrm{T}$ (field integral of $3\,\mathrm{T\cdot m}$), and a trigger system made of four planes of resistive plate chambers performing the selection of high transverse momentum muons. Muons penetrating the whole spectrometer length are measured with a momentum resolution of about $1\div2\,\%$ and $90\,\%$ efficiency (for $p_{\mathrm{t}} > 3\,\mathrm{GeV}/c$) [16].

---

[1] Even if less than 1% of all heavy-quark pairs form quarkonium bound states.

[2] $D^0$ mesons have proper decay length $c\tau = (123.0 \pm 0.4)\,\mu\mathrm{m}$ and $B$ mesons of about 500 $\mu$m [17].

[3] TRD is not used as a tracking device in the analysis presented hereafter.

# CHARM PRODUCTION MEASUREMENT VIA HADRONIC DECAYS

ALICE capability for measuring direct charm production through the reconstruction of the exclusive hadronic decays have been evaluated in the benchmark $D^0 \to K^+\pi^-$ two body decay mode (and charge conjugate) of branching ratio $(3.83 \pm 0.09)\%$ [13,18,19]. Such a measurement provides a direct way to access charm transverse momentum spectrum which is of crucial interest when trying to assess the effects induced by the nuclear medium in p-Pb and Pb-Pb collisions. A $D^0$ candidate consists of a pair of oppositely charged tracks originating from a secondary vertex (cf. Fig. 1 left hand panel) which has a $K^+\pi^-$ mass in the range $M_{D^0} \pm 3\sigma$ *i.e.* $|\Delta M| < 12\,\mathrm{MeV}/c^2$ in the present study (cf. Fig. 1 right hand panel).

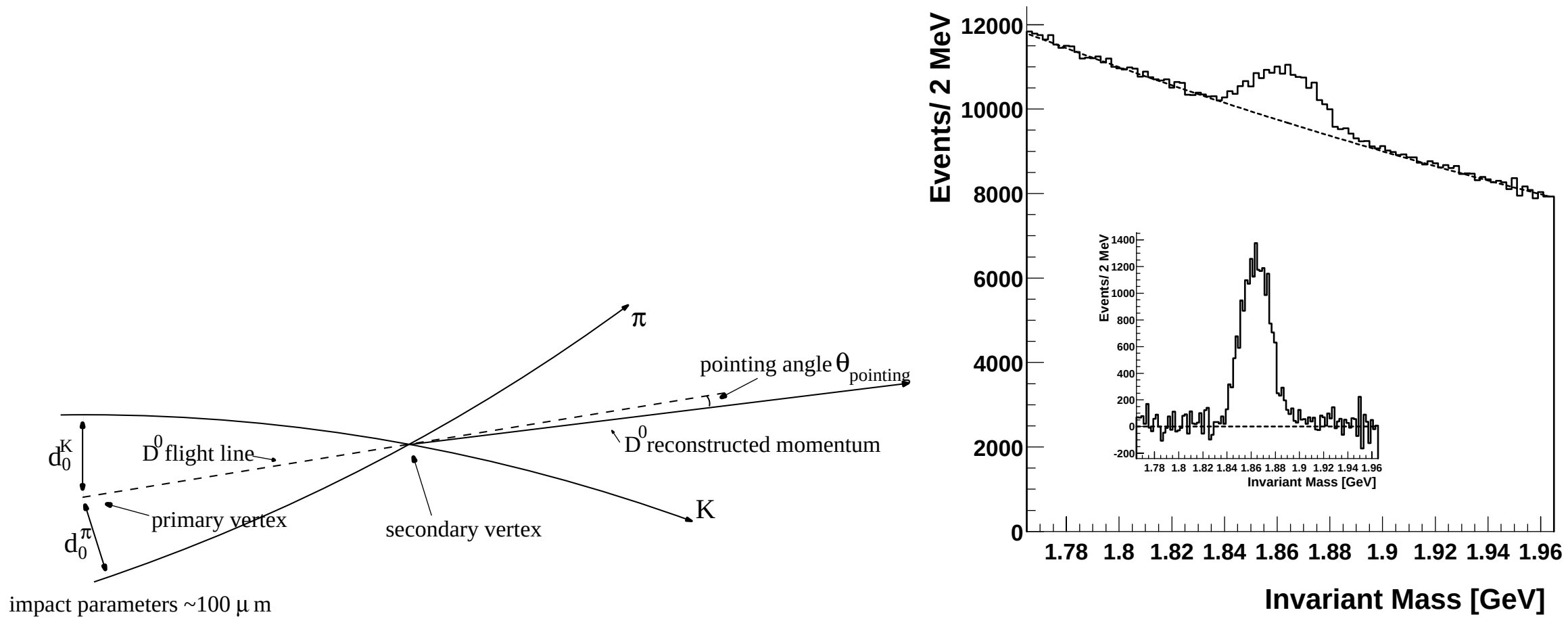

**FIGURE 1.** Schetch of the $D^0 \to K^+\pi^-$ decay (left hand panel). $K^+\pi^-$ invariant-mass distribution corresponding to $10^7$ central Pb–Pb events (right hand panel); the background-subtracted distribution is shown in the insert.

Extracting a $D^0$ signal with a good significance in central Pb-Pb collisions requires a drastic selection procedure to reduce the huge combinatorial background by at least 6÷7 orders of magnitude ($S/B \sim 10^{-6}$ in the mass range $M_{D^0} \pm 3\sigma$ before any geometrical or kinematical selection). At dedicated set of cuts (distance of closest approach between the tracks, $K$ and $\pi$ minimum transverse momentum, and decay angle) detailed in Ref. [18,19], improves the $S/B$ by about two orders of magnitude. The $S/B$ ratio is then increased by three additional orders of magnitude imposing a combined cut on the values of $d_0^K \times d_0^\pi$ and $\cos\theta_{\mathrm{pointing}}$ [4] ($d_0^K \times d_0^\pi < -40{,}000\,\mu\mathrm{m}$ and $\cos\theta_{\mathrm{pointing}} > 0.98$ in this study). Cut tuning has been performed in the present study for each separate $D^0$ transverse momentum bin. With $10^7$ Pb-Pb events, we expect a $p_{\mathrm{t}}$-integrated significance of 37 and larger than 10 up to $p_{\mathrm{t}} \sim 10$ GeV$/c$. The lower $p_{\mathrm{t}}$ limit is expected to be 1 GeV$/c$ and even lower in p-p collisions (cf. Fig. 2).

[4] $\theta_{\mathrm{pointing}}$ is defined as the angle between the $D^0$ candidate momentum and the line joining the primary with the secondary vertex as sketched in Fig. 1 on the left hand panel

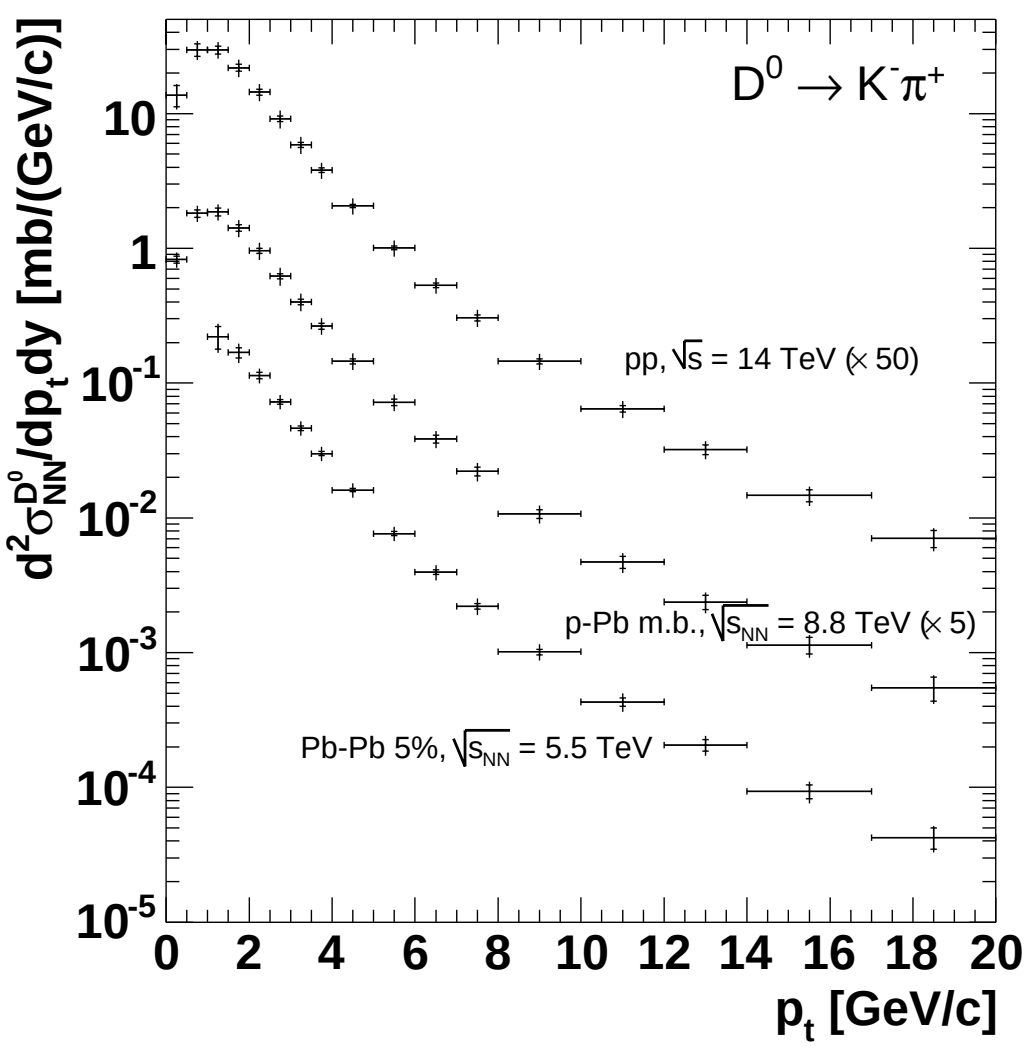

**FIGURE 2.** $p_t$-differential cross-section per nucleon-nucleon collision for $D^0$ production as expected to be measured with $10^7$ central Pb-Pb events, $10^8$ minimum-bias p-Pb events, and $10^9$ p-p minimum-bias events. Statistical (inner bars) and quadratic sum of statistical and $p_t$-dependent systematic errors (outer bars) are shown. A normalization error of 9% for Pb-Pb, 9% for p-Pb and 5% for p-p is not shown.

# BEAUTY MEASUREMENTS FROM SEMILEPTONIC DECAYS

## Single Inclusive Production Cross-Section Measurement Using Muons

Beauty production in Pb-Pb collisions will be measured in ALICE via semimuonic decays in the pseudo-rapidity region $-4 \leq \eta \leq -2.5$ [16]. Both inclusive muon and opposite-sign dimuon are considered in this study. The dimuon sample is divided into two topologically distinct contributions: $b$-chain decays (named $BD_{\rm same}$ in Fig. 3(d)) of low mass $M_{\mu^+\mu^-} < 5\,{\rm GeV}/c^2$ and high transverse momentum (cf. Fig. 3(a)) and muon pairs where the two muons originate from different quarks ($BB_{\rm diff}$) emitted at large angles resulting in large invariant masses $M_{\mu^+\mu^-} > 5\,{\rm GeV}/c^2$ (cf. Fig. 3(b)). The beauty signal is enhanced with respect to other sources (charm and $\pi/K$ decay-in-flight) applying a low $p_t$ cut-off set to 1.5 GeV/$c$ in this study. Using Monte Carlo predicted line shapes for $c\overline{c}$, $b\overline{b}$, and decay background, fits are carried out to find the $b\overline{b}$ fraction in the different data sets as presented in Fig. 3(a), (b), and (c). Finally muon level cross-sections are converted into inclusive $b$-hadron cross-section following the method initially developped for the UA1 experiment [19]. The expected performance for the measurement of $b$-hadron cross-section for $10^7$ central Pb-Pb collisions is plotted in Fig. 3(d). $b$-decay muon statistics is large over the whole $p_t$ range allowing a tight mapping of the production cross-section up to $p_t \sim 20\,{\rm GeV}/c$.

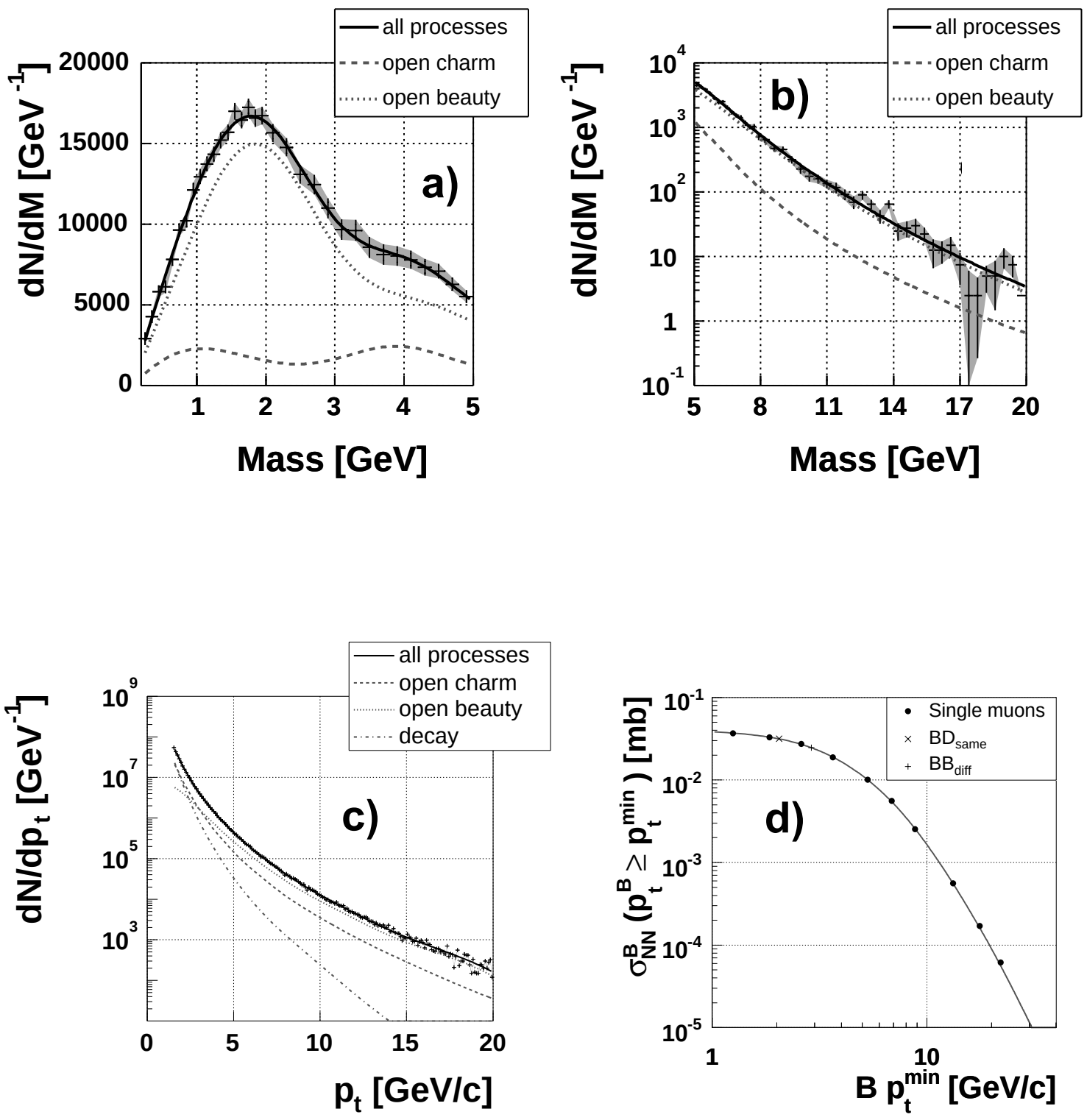

**FIGURE 3.** Background subtracted invariant mass distributions of $\mu^+\mu^-$ pairs produced in $10^7$ central Pb-Pb collisions in the low (a) and high mass regions (b). A $p_t > 1.5$ GeV/$c$ cut-off has been applied to muon tracks. Charm and beauty signals are plotted in dashed and dotted line respectively. (c) Single muon transverse momentum distribution. (d) Expected performance for the measurement of the inclusive $b$-hadron cross-section in $-4 < y^B < -2.5$ as a function of $p_t^{min}$. PYTHIA predictions (solid line) used to produce the signal are also shown.

## Single Inclusive Production Cross-Section Measurement Using Electrons

Electrons from semileptonic decay of $b$-quarks are characterized by a hard transverse momentum spectrum and a large average impact parameter $(\langle d_0 \rangle \simeq 300\ \mu\text{m})$ with respect to other electron sources: pions misidentified as electrons, decays of primary prompt charmed hadrons, decays of light mesons (*e.g.* $\pi^0$ Dalitz, $\rho$, $\omega$, $K$), and photon conversions in the beam pipe or in the ITS inner layers (cf Fig. 4). The ALICE detection strategy will then rely on the selection of displaced tracks identified as electrons.

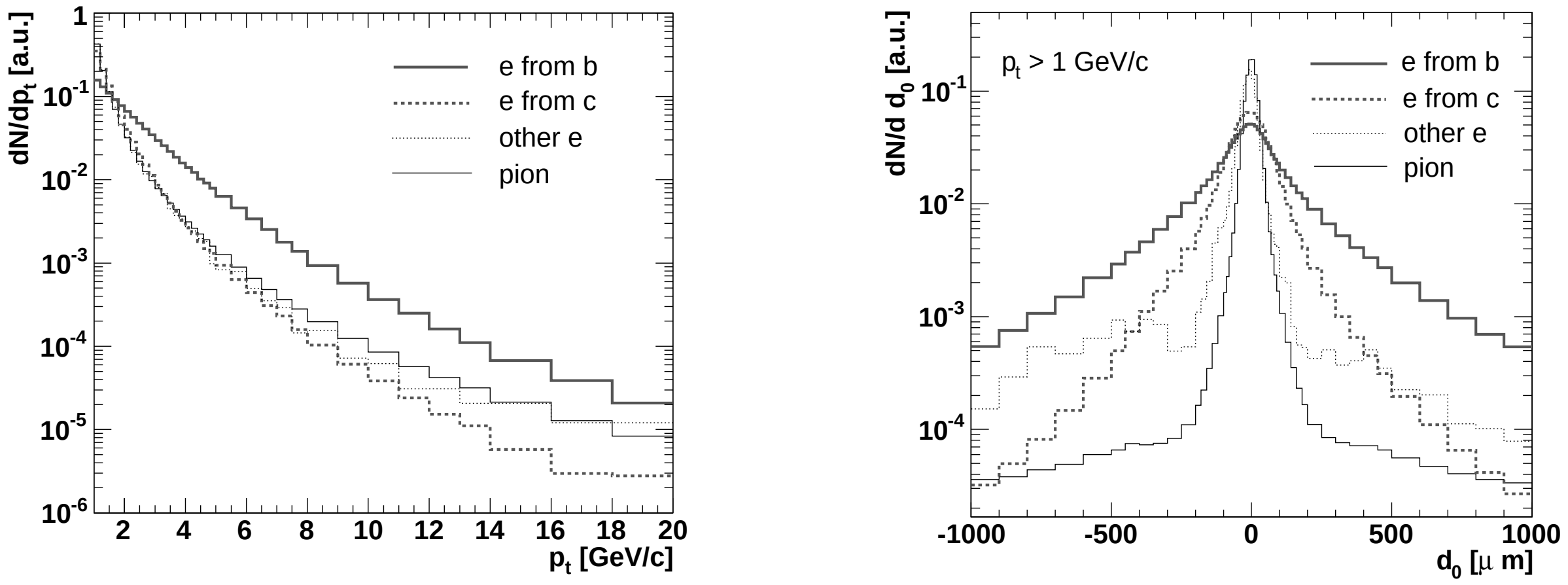

**FIGURE 4.** Comparison of electron transverse momentum (left hand panel) and impact parameter projection in the transverse plane (right hand panel) from beauty, charm, light meson, conversion, and pion decays (including a 1 GeV/$c$ $p_{\rm t}$ cut-off).

Electron identification is performed combining information from TRD and TPC. For a TRD electron identification efficiency of $\simeq 90\,\%$, a $1\,\%$ pion contamination is expected. This efficiency is expected to be roughly constant in the momentum range containing the largest beauty fraction $(1 \div 6\,\mathrm{GeV}\,/\,c)$. Electrons crossing the TPC are separated from heavier particles by their specific energy loss $\mathrm{d}E\,/\,\mathrm{d}x$. Considering only tracks tagged as electrons in the TRD, the contamination from charged kaons and protons is expected to be negligible, and electron-tagging probability for electrons is expected to be $\simeq 90\,\%$ while electron-tagging probability for pions is expected to be at the level of 1 % up to $p \sim 2 \div 3\,\mathrm{GeV}\,/\,c$ (increasing as a function of $p$ to exceed 60 % for $p \sim 15\,\mathrm{GeV}\,/\,c$). TRD-TPC electron identification reduces pion contamination by 4 orders of magnitude.

The expected beauty signal purity and statistics as a function of the impact parameter cut-off for different values of the $p_{\rm t}$ threshold are shown in Fig. 5 left and right hand panel respectively [21]. For instance, $p_{\rm t} > 2$ GeV/c and $200 < |d_0| < 600$ $\mu$m are expected to provide, for $10^7$ central Pb-Pb collisions, an electron sample of $8{\times}10^4$ with a 90% purity. An upper limit on $d_0$ is applied here in order to reduce long lived strange particles and tracks suffering from large angle scatterings in detector materials.

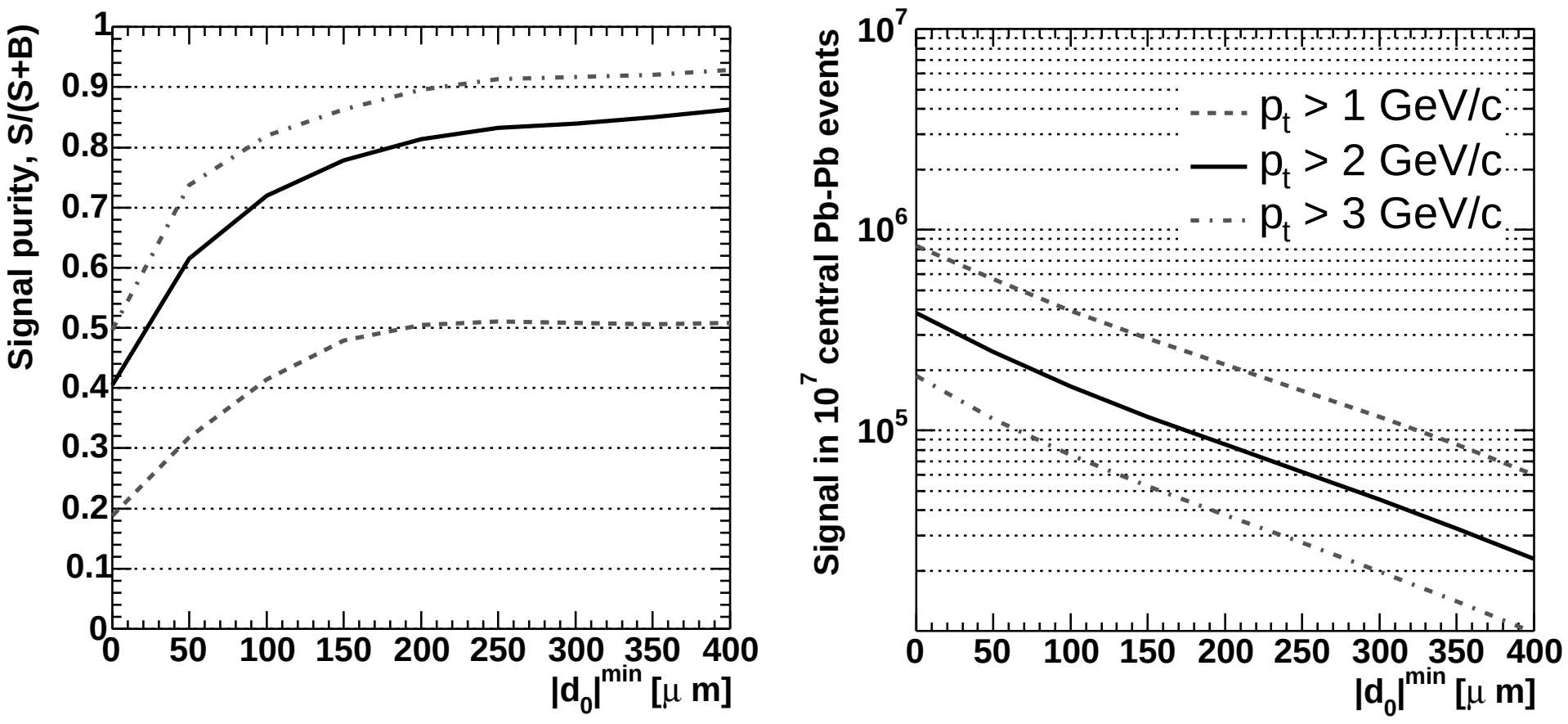

**FIGURE 5.** Expected signal purity (left hand panel) and signal (right hand panel) of reconstructed beauty decay electrons as a function of the impact parameter threshold for three different values of transverse momentum cut-off in $10^7$ central Pb-Pb events.

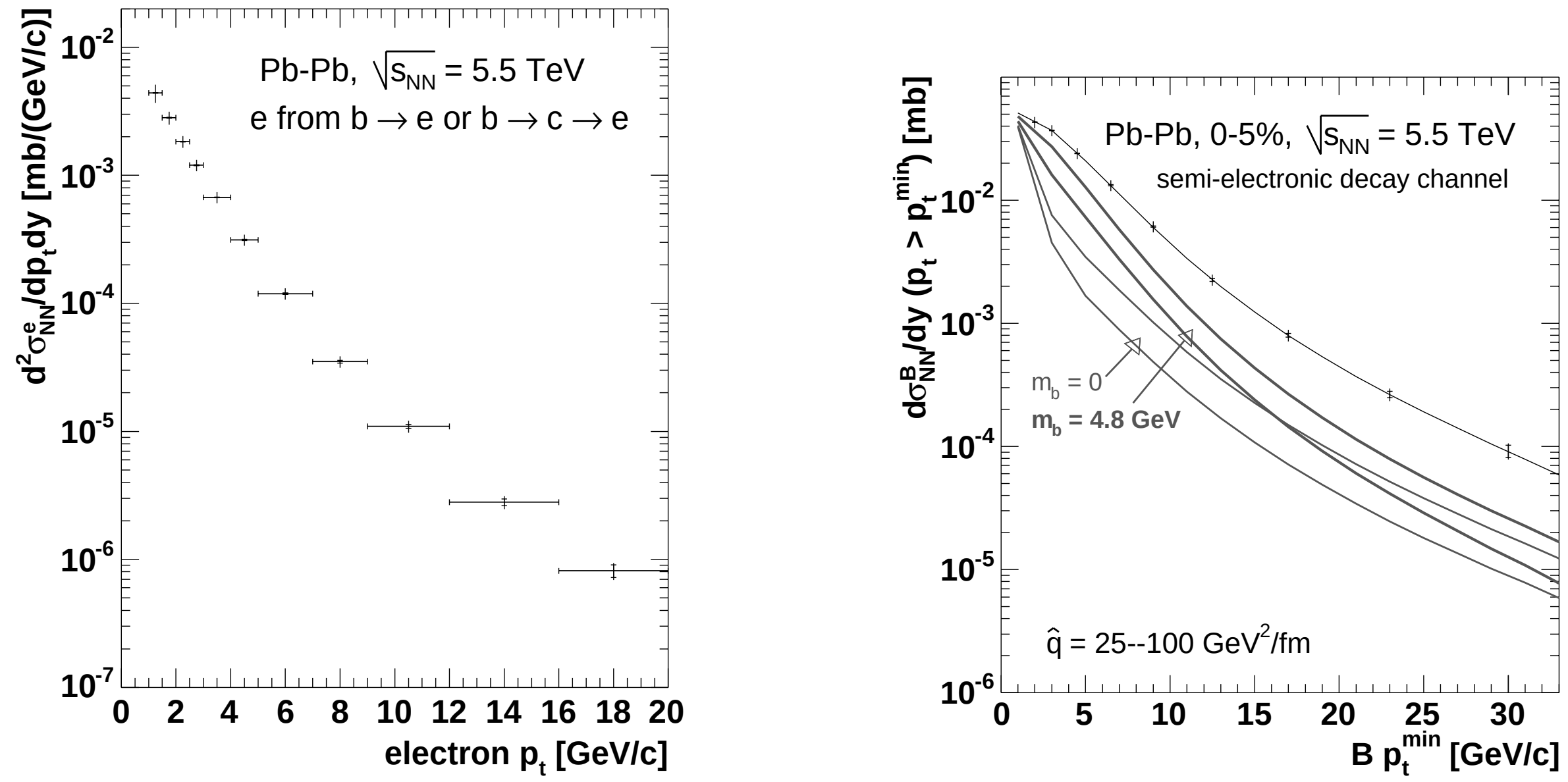

**FIGURE 6.** Expected performance for the measurement of the doubly differential *b*-decay electron cross-section per nucleon-nucleon collision as a function of transverse momentum for $10^7$ central Pb-Pb collisions (left hand panel). Statistical (inner bars) and systematic (outer bars) errors are displayed. Corresponding inclusive *b* -hadron cross-section as a function of $p_t^{min}$ (right hand panel). An overall normalization error (9%) is not included. Quenching predictions from calculation [7] are superimposed for a massive and massless *b* quark and a transport coefficient $\hat{q}$ in the range $25 \div 100\ \mathrm{GeV}^2 / c$ .

Finally, the *b*-decay electron transverse momentum distribution is obtained after subtraction of the remaining background (cf. Fig. 6 left hand panel). We intend to subtract the remaining background from charm using the hadronic charm production cross-section measurement (described earlier) while the steeply falling ($< 2\,\%$ for $p_t \gtrsim 4$ GeV/$c$) decay background (from pions tagged as electrons and other sources) will be estimated with Monte Carlo simulations from the measured charged pion $p_t$ distributions. The corresponding *b*-hadron cross-section per unit of rapidity as a function of the minimum transverse momentum ( $p_t^{min}$ ) is inferred "à la" UA1 just like for the muon channel as described in the previous section and the expected measurement shown in Fig. 6 (right hand panel). *b*-quark quenching effects on the *b*-hadron cross-section are also represented in Fig. 6 (right hand panel) but only with a view of illustration since any suppression pattern will be assessed from a comparison with the reference cross-section provided by p-p data.

## CONCLUSIONS AND OUTLOOK

With its excellent tracking, vertexing, and particle identification capabilities, ALICE has promising perspectives for heavy flavour measurements with abilities to: fully reconstruct hadronic charm decay topologies and address semileptonic *b*-quark decays in p-p, p-Pb and Pb-Pb collisions with relatively low transverse momentum thresholds and large rapidity coverage. Heavy flavour transverse momentum spectra are of special interest when studying the properties of dense QCD matter created in heavy ion collisions where new effects are expected as compared to elementary nucleon-nucleon interactions (shadowing, in-medium quenching ...). Heavy quark transverse momentum distribution measurements from large statistics data samples ($10^7$ central Pb-Pb collisions) have been presented. Channels discussed in these proceedings will be supplemented, as far as heavy flavour production is concerned, by measuring other observables (dilepton correlations, secondary J/$\psi$ from $b$ decay, and $b$-tagged jets...). The assessment of the performance of these new physics channels is currently underway.

## ACKNOWLEDGMENTS

Part of this work was supported by the EU Integrated Infrastructure Initiative Hadron-Physics Project under contract RII3-CT-2004-506078.